 \documentclass[final,5p,times]{elsarticle}

\usepackage{graphics}
\usepackage{graphicx}
\usepackage{epsfig}

\usepackage{amssymb}
\usepackage{amsthm}

\biboptions{sort&compress}

\newcommand{\od}{\mathrm{d}}
\newcommand{\pd}{\partial}
\usepackage{enumerate,amsmath,bm,amssymb}
\usepackage{color}

\journal{Computers \& Fluids}

\begin{document}

\begin{frontmatter}



\title{Analysis of electro-osmotic flow in a
microchannel with undulated surfaces}

\author[a1,a2]{Hiroaki Yoshida\corref{cor1}}
\ead{h-yoshida@mosk.tytlabs.co.jp}
\author[a1,a2]{Tomoyuki Kinjo}
\author[a1,a2]{Hitoshi Washizu}
%

\address[a1]{Toyota Central R\&D Labs., Inc., Nagakute, Aichi 480-1192, Japan}
\address[a2]{Elements Strategy Initiative for Catalysts and Batteries
 (ESICB), Kyoto University,
Kyoto 615-8245, Japan}

\cortext[cor1]{Corresponding author.}





\begin{abstract}
The electro-osmotic flow through a channel between two undulated
surfaces induced by an external electric field is investigated. 
The gap of the channel is very small and comparable
to the thickness of the electrical double layers.
A lattice Boltzmann simulation is carried out
on the model consisting of the Poisson
equation for electrical potential, the Nernst--Planck equation for
ion concentration, and the Navier--Stokes {\color{black}equations} for flows of
the electrolyte solution.
An analytical model that predicts the flow rate is also derived under
the assumption that the channel width is very small compared with the
characteristic length of the variation along the channel.
The analytical results are compared with the numerical results obtained by using the
lattice Boltzmann method.
In the case of a constant surface charge density along the channel,
the variation of the channel width 
reduces the electro-osmotic flow,
and the flow rate is smaller than that of a straight channel. 
In the case of a surface charge density 
distributed inhomogeneously,
one-way flow occurs
even under the restriction of a zero net surface charge along the channel.
\end{abstract}

\begin{keyword}
Electro-osmotic flow\sep
Electrical double layer\sep
Lattice Boltzmann method\sep
Lubrication approximation theory


\end{keyword}

\end{frontmatter}



%
%
\section{\label{sec_intro}Introduction}
Adjacent to the interface between
an electrolyte solution and a charged solid surface,
an electrical double layer is formed.
The thickness of the double layer ranges from a few nanometers to hundreds
of nanometers, depending on the salt concentration.
Since the ion concentration in the electrical double layer is highly inhomogeneous
and the local charge neutrality is broken, 
interaction with an externally applied electric field
can cause a driving force acting on the electrolyte solution.
The driving force is the main factor
for electrokinetic effects,
such as the electro-migration of colloid particles
and the electro-osmotic flow inside
microchannels~\cite{Israelachvili2011,KBA2005},
which are especially conspicuous in small-scale systems.

Micro- and nano-fabrication techniques have developed greatly in
recent years, and so researchers have been able
to control and use electrokinetic phenomena 
in engineering applications.
For example, an electro-osmotic pump without moving parts~\cite{UTL+2006,YW2013} 
and an energy-harvesting device using the driving force
induced by the salt-concentration gradient~\cite{SPB+2013} have been proposed very recently.
Along with the expectations for engineering applications,
fundamental research relevant to
electrokinetic phenomena in small-scale systems,
ranging from nano- to micrometers,
has also {\color{black}attracted attention~\cite{CGS2007,SHR2008,BC2010B,ZY2012,DGC2013,ZRA2014,YMK+2014,YMK+2014a,MNM2014,PPD+2014}.}
Particularly, the advances in observing and processing techniques
have prompted research
focusing on surface properties such as {\color{black}roughness and structure~\cite{TS2006,WWC2007,WK2009,XMS+2009,BN2010,MS2010,LWCR2010,BB2015}.}

In the present study, 
to clarify the effects of surface properties
on electrokinetic phenomena,
we investigate electro-osmotic flows
between two surfaces, which have a periodic structure
and a non-uniform charge distribution,
by means of both numerical and analytical approaches.
The numerical analysis is based on the 
coupled lattice Boltzmann method for solving the 
Navier--Stokes {\color{black}equations}, the Nernst--Plank equation,
and the Poisson equation~\cite{YKW2014}.
The numerical solutions to these equations, i.e., the electrolyte flow, the ion
concentrations, and the electrical field, are directly obtained.
{\color{black}In the analytical approach,
the lubrication approximation theory~\cite{Ghosal2002,NZ2012,NQ2014}
is applied to the system with electrical double layers of finite thickness,
to derive a model equation
 that predicts the flow rate under
the assumption of moderate variation of the surface structure along the channel.}
With these approaches, we investigate the electro-osmotic flow in
microchannels formed by the surfaces that are undulated and
(a) charged negatively at
a constant surface charge density or
(b) charged non-uniformly along the channel such that the net surface charge vanishes.
For both cases, the electro-osmotic flow rate is evaluated and 
the dependency on geometrical parameters,
such as the amplitude of the surface shape, is discussed.

%
%
%

\section{\label{sec_problem}Problem and basic equations}

%
\subsection{\label{ssec_problem}Channel with undulated surfaces}
\begin{figure}[tb]
%
\begin{center}
\vspace*{0mm}
\includegraphics[scale=0.7]{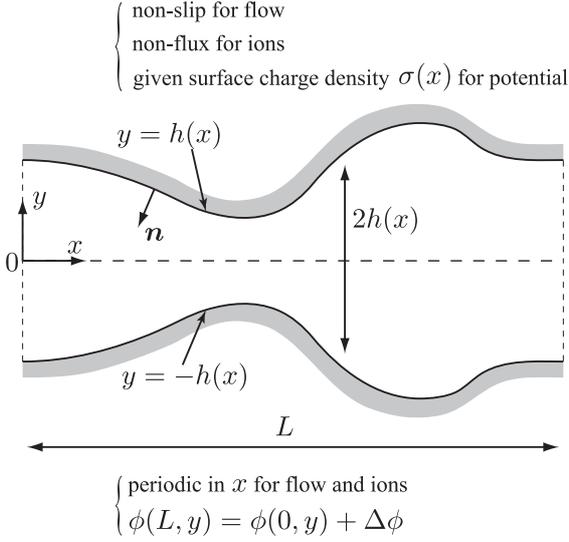}
\end{center}
%
\caption{Geometry of the problem.}
\label{fig01}
\end{figure}

Let us consider
a channel between two walls,
each of which has a periodic structure of period $L$
in the $x$ direction. The positions of the interfaces
are expressed as $y=\pm h(x)$ (Fig.~\ref{fig01}).
The surface charge density on the channel walls is assumed to be 
a given function $\sigma(x)$.
The two-dimensional domain between the surfaces ($-h(x)<y<h(x)$)
is filled with a $1:1$ electrolyte solution,
and the electrical double layer is formed near the
interfaces. 
We investigate the electro-osmotic flow
caused by an electrical field applied in the $x$ direction
by applying the governing equations described in the next subsection.

%

%
\subsection{\label{ssec_equation}Governing equations}

We assume a Newtonian fluid for the electrolyte solution,
and then the flow is described by the Navier--Stokes {\color{black}equations}:
\begin{align}
&\dfrac{\pd u_j}{\pd x_j}=0,
\label{s2-div}\\
&\dfrac{\pd u_i}{\pd t}+u_j\dfrac{\pd u_i}{\pd
 x_j}=-\dfrac{1}{\rho_0}\dfrac{\pd p}{\pd x_i}
+\nu\dfrac{\pd^2 u_i}{\pd x_j^2}+\dfrac{F_i}{\rho_0},
\label{s2-ns}
\end{align}
where $t$ is the time and $\bm{x}$ is the spatial coordinate. 
In the present paper, we use either boldface letters or assign
indexes $i$ and $j$ to designate vector elements. The summation
convention is assumed for repeated indexes.
The functions $\bm{u}(t,\bm{x})$ and $p(t,\bm{x})$ are the flow velocity
and the pressure of the electrolyte solution, respectively,
and $\bm{F}(t,\bm{x})$ is the body force per unit volume. 
The density $\rho_0$ and the {\color{black}kinematic viscosity} $\nu$ of the
electrolyte solution are assumed to be constant.

The mass conservation equation for the ion species is written as
\begin{align}
&\dfrac{\pd C_{m}}{\pd t}+\dfrac{\pd J_{mj}}{\pd x_j}=0,
\label{s2-npc}\\
&J_{mi}=-\dfrac{ez_{m}D_{m}}{k_{\mathrm{B}}T}C_{m}\dfrac{\pd \phi}{\pd x_i}-D_{m}\dfrac{\pd
 C_{m}}{\pd x_i}+C_{m}u_i,
\label{s2-np}
\end{align}
where $C_{m}$ and $\bm{J}_{m}$ denote the concentration and the flux
of an ion, respectively, with 
$m=\textrm{a}$ for the anion and $m=\textrm{c}$ for the cation.
Here, $e$ is the unit charge, $k_{\mathrm{B}}$ is Boltzmann's constant,
and $T$ is the temperature. 
The constants $z_m$ and $D_m$ are the valence and
the diffusion coefficient of species $m$, respectively.
Equation~\eqref{s2-np} is referred to as the Nernst--Planck model \cite{N1991},
where the first, second, and third terms on the right-hand side of the flux equation are
the contributions of electrochemical migration, diffusion, and
convection of the electrolyte solution, respectively.

Finally, the electrical potential $\phi$ is governed by the following Poisson equation:
\begin{equation}
\varepsilon\dfrac{\pd^2
 \phi}{\pd
 x_j^2}=-\rho_{\mathrm{e}},
\label{s2-poisson}
\end{equation}
where $\varepsilon$ is 
the dielectric constant of the electrolyte solution.
The local charge density $\rho_{\mathrm{e}}$ is 
defined in terms of the ion concentration as
\begin{equation}
\rho_{\mathrm{e}}=\sum_m Fz_mC_m,
\label{s2-charge}
\end{equation}
where $F$ is Faraday's constant.
With this local charge density, the body force $\bm{F}$ in Eq.~\eqref{s2-ns}
is defined as the interaction with the electric field:
\begin{equation}
F_i=-\rho_{\mathrm{e}}\dfrac{\pd\phi}{\pd x_i}.
\label{s2-force}
\end{equation}

{\color{black}
If the convection term in Eq.~\eqref{s2-np} is negligible compared with 
the other two terms and
a unique value of potential is defined at $C_m=C_0$, 
integration of Eq.~\eqref{s2-np} yields the Boltzmann distribution
$C_m=C_0\exp(-ez_m\phi/k_{\mathrm{B}}T)$.
With this formula, Eq.~\eqref{s2-poisson} reduces to
the Poisson--Boltzmann equation.
In the electrokinetic flows considered in the present paper, however,
although the convection term is sufficiently small,
specifying a unique potential value at $C_m=C_0$ is difficult
because of the external potential gradient.
We therefore apply the original set of equations described here
in the numerical simulations in Section~\ref{sec_result}.
Then the ion distribution affected by the external potential
is obtained as shown in Section~\ref{ssec_distributed},
which the Poisson--Boltzmann equation decoupled from the external
potential field fails to capture.
}

%
\subsection{\label{ssec_bc}Boundary conditions on solid-liquid interface}

For the flow velocity,
the ordinary non-slip condition is assumed
at the solid-liquid interface:

\begin{equation}
u_i=0, \qquad \textrm{at}\ y=\pm h(x).
\label{s2-nonslip}
\end{equation}
Note that
in nano-scale flows, 
the simple non-slip condition is not sufficient
and so a model describing the slip taking place at the interface
is necessary.
However, since the scale of the problems considered in the present paper is
relatively large (at a scale of micrometers),
we assume the non-slip condition to be valid.
For the ion concentration,
no flux goes across the boundary, which is formulated
simply as 
\begin{equation}
n_jJ_j=0,\qquad \textrm{at}\ y=\pm h(x),
\label{s2-nonflux}
\end{equation}
where $\bm{n}$ is the unit normal vector pointing inward to the fluid
 region.
If we substitute Eq.~\eqref{s2-np} into Eq.~\eqref{s2-nonflux}, then we have
\begin{equation}
-\dfrac{ez_{m}D_{m}}{k_{\mathrm{B}}T}C_{m}n_j\dfrac{\pd \phi}{\pd x_j}
-D_{m}n_j\dfrac{\pd
 C_{m}}{\pd x_j}
+C_{m}n_ju_j=0.
\label{s2-nonflux2}
\end{equation}
This form of the Neumann-type boundary condition seems rather complex to
implement in the lattice Boltzmann method.
However, with the scheme for the Nernst--Planck model described in the next section
we can impose this condition in a simple manner.
Finally, the boundary conditions for the electrical potential
are given as the following Neumann-type condition:
\begin{equation}
-\varepsilon n_j\dfrac{\pd \phi}{\pd x_j}=\sigma,
\qquad \textrm{at}\ y=\pm h(x). 
\label{s2-sfcharge}
\end{equation}
%

%
%
%
\section{\label{sec_lbm}Lattice Boltzmann method}

%
\subsection{\label{ssec_lbe}Lattice Boltzmann equation}
In this subsection, we outline 
the numerical method
based on the lattice Boltzmann method (LBM)~\cite{MZ1988,QDL1992,CD1998,S2001,LL2000}
for solving the following set of model equations for electrokinetic flows:
(I) the Poisson equation \eqref{s2-poisson} with Eq.~\eqref{s2-charge},
(II) the Nernst--Planck equation \eqref{s2-npc} with Eq.~\eqref{s2-np},
and (III) the Navier--Stokes {\color{black}equations} \eqref{s2-ns} with Eqs.~\eqref{s2-div}
and \eqref{s2-force}.

The four lattice Boltzmann equations are assigned 
to the unknown variables $\phi$, $C_{\mathrm{a}}$, $C_{\mathrm{c}}$, and $\bm{u}$.
First, the lattice Boltzmann algorithm generally used
for (I) through (III) is outlined, and 
then we comment on the scheme for the Nernst--Planck equation.
The complete description of the framework including the rules for the boundary conditions
is found in Ref.~\cite{YKW2014}.

The LBM tracks
the behavior of the distribution function 
$f_{\alpha}(t,\bm{x})$, where $\alpha=0,1,2,$ $\ldots$, $N$,
rather than that of 
the unknown variable for the target partial differential equations.
Here and in what follows, the subscript $\alpha$ is used to indicate the quantities
corresponding to the directions of the discrete velocities, such as $f_{\alpha}$ above.
The values of the distribution function travel
over a regular spatial lattice with the assigned discrete velocities, 
of which the direction is defined in terms of the vector $\bm{e}_{\alpha}$.
The explicit expression of the vector $\bm{e}_{\alpha}$ depends on the type
of partial differential equation to be solved.

{\color{black}
Since we have four unknown variables,
namely the electrical potential, the concentration of cation and anion,
and the flow velocity,
we use four distribution functions, denoted by $f_{\alpha}^{s}$
with $s=\mathrm{p}$, $\mathrm{c}$, $\mathrm{a}$, and $\mathrm{u}$.
The unknown variables
are calculated as moments of the distribution function:
\begin{align}
&\phi=\sum_{\alpha}f_{\alpha}^{\mathrm{p}},
\label{s3-def_p}\\
&C_{\mathrm{c}}=\sum_{\alpha}f_{\alpha}^{\mathrm{c}},\quad
C_{\mathrm{a}}=\sum_{\alpha}f_{\alpha}^{\mathrm{a}},
\label{s3-def_c}\\
&u_i=\dfrac{\Delta x}{\rho\Delta t}\sum_{\alpha}e_{\alpha i}f_{\alpha}^{\mathrm{u}},\quad
\textrm{with}\ \rho=\sum_{\alpha}f_{\alpha}^{\mathrm{u}},
\label{s3-def_u}
\end{align}
where $\Delta t$ and $\Delta x$
are the time step and the grid interval, respectively.
}

The lattice Boltzmann equation used in the present study is written as 
{\color{black}
\begin{align}
&f_{\alpha}^{s}(t+\Delta t,\,\bm{x}+\bm{e}_{\alpha}\Delta x) - f_{\alpha}^{s}(t,\bm{x})
=\dfrac{1}{\tau^s}(f^{s(\mathrm{eq})}_{\alpha}-f_{\alpha}^s)(t,\bm{x})
\nonumber\\
& \hspace*{3cm}+\dfrac{\Delta t}{2}
\bigg(g^{s}_{\alpha}(t,\bm{x})+g^{s}_{\alpha}(t,\,\bm{x}+\bm{e}_{\alpha}\Delta x)\bigg),
\label{s3-lbe}
\end{align}
where the function $g^{\mathrm{u}}_{\alpha}$ corresponds to the forcing term
for solving the Navier--Stokes {\color{black}equations}, and $g^{\mathrm{p}}_{\alpha}$ corresponds
to the source term for solving the Poisson equation,
while $g^{\mathrm{c}}_{\alpha}=g^{\mathrm{a}}_{\alpha}=0$.}
{\color{black}Note that, in our formulation, $\bm{e}_{\alpha}$ is defined as
a dimensionless vector designating the directions of the discrete
velocities~\cite{YKW2014}.}
The first term on the right-hand side is
the collision term, which defines the relaxation process during a time step.
The coefficient $\tau^s$ defines the
relaxation time and is related to the dielectric constant, the diffusion coefficients,
and the viscosity~\cite{YKW2014}:
{\color{black}
\begin{align}
\tau^{\mathrm{p}}
=\dfrac{1}{2}+\dfrac{\Delta t}{\Delta
 x^2\Lambda^{\mathrm{p}}}\varepsilon,
\quad
\tau^{\mathrm{c,a}}
=\dfrac{1}{2}+\dfrac{\Delta t}{\Delta
 x^2\Lambda^{\mathrm{c,a}}}D_{\mathrm{c,a}},
\quad
\tau^{\mathrm{u}}
=\dfrac{1}{2}+\dfrac{\Delta t}{\Delta
 x^2\Lambda^{\mathrm{u}}}\nu,
\label{s3-def_tau}
\end{align}
where the value of constant $\Lambda^s$ depends on the discrete velocity set $\bm{e}_{\alpha}$~\cite{YKW2014}.}
The equilibrium distribution function $f^{s(\mathrm{eq})}_{\alpha}$ is
expressed in terms of the local value of the 
physical quantities, such as
the electrical potential, the ion concentration, and the flow velocity. 
The collision term used herein is the 
single-relaxation-time (SRT) method, in which 
a common value of the relaxation-time coefficient $\tau^s$ is assigned to 
all directions of $\alpha$. 
{\color{black}
Although 
various types of collision operator
have been presented to increase stability and/or
accuracy~\cite{LL2000,Hetal2002,AKO2003,GGK2006,Ginzburg2012},
}
we prefer the simplicity of the SRT for the present study,
because the geometrical setup and the parameters used in Section~\ref{sec_result} are
not very severe.
The extension of the present numerical framework 
to the one using the multiple-relaxation-time collision operator,
which exhibits robustness for severe parameter set,
is straightforward, if the techniques described in
Refs.~\cite{LL2000,Hetal2002,GZS2008,YN2010} are applied.
The complete definitions of
$\bm{e}_{\alpha}$, $f^{s(\mathrm{eq})}_{\alpha}$ and $g^s_{\alpha}$ are found in Ref.~\cite{YKW2014}.

To implement the LBM, 
Eq.~\eqref{s3-lbe} is split into the
collision process and the streaming process,
and the distribution function is updated in an explicit manner:

\vspace*{3mm}
\noindent \underline{Collision}:
\vspace*{-3mm}
\begin{equation}
{\color{black}
\hat{f}^s_{\alpha}(t,\bm{x})=f^s_{\alpha}(t,\bm{x})+
\dfrac{1}{\tau^s}(f^{s(\mathrm{eq})}_{\alpha}-f_{\alpha}^s)(t,\bm{x})
+\dfrac{\Delta t}{2}g_{\alpha}^s(t,\bm{x}),}
\label{s3-proc-collision}
\end{equation}

\noindent \underline{Streaming}:
\vspace*{-3mm}
\begin{equation}
{\color{black}
f_{\alpha}^s(t+\Delta t,\bm{x}+\bm{e}_{\alpha}\Delta
 x)=\hat{f}_{\alpha}^s(t,\bm{x})
+\dfrac{\Delta t}{2}g_{\alpha}^s(t,\bm{x}+\bm{e}_{\alpha}\Delta x).}
\label{s3-proc-stream}
\end{equation}
%

{\color{black}
In solving the Navier--Stokes {\color{black}equations},
the simple halfway bounce-back rule 
for the non-slip boundary condition
is applied 
regarding the
lattice points outside the channel as the solid phase.
Similarly, 
the halfway bounce-back rule
for the non-flux boundary condition is utilized
in solving the Nernst--Planck equation.
On the other hand, a modified boundary rule for the curved Neumann
boundary condition is 
applied for solving the Poisson equation.
This is because the 
simple bounce-back rule causes serious errors 
when implementing the curved Neumann boundary condition with
inhomogeneous term (Eq.~\eqref{s2-sfcharge}),
and special treatment is necessary, as demonstrated in
Refs.~\cite{YN2010,GH2013,LMK2013}.
In the present study, the scheme utilizing the level set method
described in detail in Ref.~\cite{YN2010} is employed, because the implementation
is rather simple once the level set function describing the surface
shape is constructed.
}

{\color{black}
Strictly speaking, the numerical solution $\bm{u}$ obtained with the LBM 
includes 
an artificial compressibility,
and different formulations have been proposed to
deal with this matter (e.g., Ref.~\cite{HL1997C}).
We employ, however,  the most widely used formulation 
for the equilibrium distribution function (e.g., Refs.~\cite{CD1998,YKW2014})
suffering from the artificial compressibility.
Since the variation of $\rho$ (or $p$) is 
controlled to remain small 
(by setting small Mach number $|\bm{u}|\Delta t/\Delta x$) so that
the incompressible Navier--Stokes {\color{black}equations} are well approximated~\cite{QO1993}, we
safely apply this method to obtain the flow field in the present analysis. 
}

We further comment on
the scheme used for solving the Nernst--Planck equation:
The electrochemical migration term 
(the first term on the right-hand side of Eq.~\eqref{s2-np})
is often regarded as the source term \cite{WK2010}, i.e.,
the function $g^s_{\alpha}$ is defined to include
the migration effect.
Although this is a straightforward strategy of dealing with this term in
the LBM, it causes difficulty in implementing
the Neumann-type boundary condition given in Eq.~\eqref{s2-nonflux2} directly.
To circumvent this difficulty,
we use the equilibrium distribution function $f^{s(\mathrm{eq})}_{\alpha}$
rather than the source term
to incorporate the electrical migration.
In the description of $\bm{J}_m$ in Eq.~\eqref{s2-np},
we are able to regard the coefficient of $C_m$ in the migration term
as a part of the convection velocity of the general convection-diffusion
equation. Accordingly, the equilibrium distribution function of the 
LBM for the general convection-diffusion equation described in
Ref.~\cite{YN2010} is defined as
\begin{align}
{\color{black}
f^{s(\mathrm{eq})}_{\alpha}=\bigg[\omega^s_{\alpha}
+\dfrac{\Delta t}{\Delta x \Lambda^{s}}
\left(u_j-\dfrac{ez_mD_m}{k_{\mathrm{B}}T}\dfrac{\pd \phi}{\pd
 x_j}\right)e_{\alpha j}\omega^s_{\alpha}\bigg]C_m,}
\label{s3-con-equi}
\end{align}
where $s=m$\,($=\mathrm{c}$ and $\mathrm{a}$) and 
$\omega^s_{\alpha}$ is the weight coefficient
and $\Lambda^s$ is defined as
$\sum_{\alpha}\omega^s_{\alpha}e_{i\alpha}e_{j\alpha}=\Lambda^s\delta_{ij}$
with $\delta_{ij}$ being Kronecker's delta.
The actual values for $\omega^s_{\alpha}$ are given in Ref.~\cite{YKW2014}.
Since the value of $\pd\phi/\pd\bm{x}$ 
is evaluated locally \cite{YKW2014,YN2010},
the communication with the 
surrounding grid points is achieved only through the streaming process (Eq.~\eqref{s3-proc-stream}).
This locality enables us to
implement the original boundary condition for the ion flux given in
Eq.~\eqref{s2-nonflux2}
in a natural form by using the standard bounce-back rules.
The artificial flux in time-dependent problems observed in using the
source-term scheme is also properly eliminated. (See Ref.~\cite{YKW2014} for a comparison with a previous
source-term scheme.) 
Among the LBMs developed for 
similar systems~\cite{CPF2004,PCF2005,GZS2005a,CS2007,WWC2007,WWL2008,WK2010,Zhang2011,ZRA2014},
we prefer to use the present algorithm because of this simple treatment of the
Nernst--Planck equation.

\begin{figure}[tb]
%
\begin{center}
\vspace*{0mm}
\includegraphics[scale=0.7]{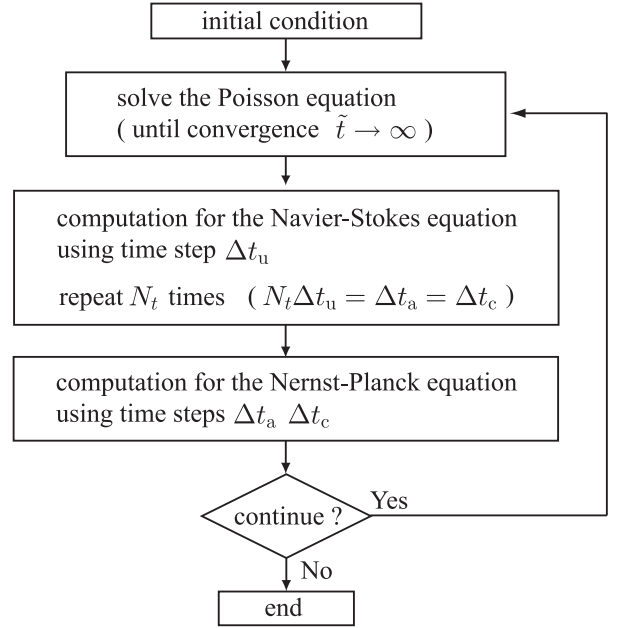}
\end{center}
%
\caption{Computational procedure for the coupled LBM.}
\label{fig02}
\end{figure}  
%

%
\subsection{\label{ssec_process}Coupling procedure}

The set of model equations includes
the Poisson equation independent of time,
in which the numerical solution $\phi$
is obtained as the long-time limit of the time-dependent solution
of the diffusion equation.
We thus introduce an artificial time axis $\tilde{t}$ for
the iteration process for solving the Poisson equation.
On the other hand, 
the ion concentrations ($C_{\mathrm{a}}$ and $C_{\mathrm{c}}$)
and the flow velocity $\bm{u}$ of the electrolyte solution
are obtained by using the common time axis $t$. 

{\color{black}
As demonstrated in Ref.~\cite{CS2008},
in solving the Poisson equation, 
the long-time limit of the diffusion equation
can lead to an undesired solution if the
initial condition is inappropriate. 
In the present work, a linear profile in the $x$-direction corresponding to the potential bias
is used  as an initial condition for the $\tilde{t}$-loop
at the very first step in physical time $t=0$.
Then such undesired solution is not encountered in the
numerical simulations presented in Section~\ref{sec_result}.
We must note here that, as is well known,
the convergence in
reaching the steady state in $\tilde{t}$ is very slow.
For $t>0$ the potential field at the previous time step is 
used as an initial guess for the loop in $\tilde{t}$,
and then the number of time steps 
required to obtain the potential field at each time step in $t$
is greatly reduced.
The very slow convergence in $\tilde{t}$-loop
at the initial stage in $t$ would be efficiently
accelerated if one uses a promising method employing
the multigrid technique as proposed in Ref.~\cite{PPB2014}.
}

Since the timescales of each transport phenomenon are different,
we need to assign different values to the time step $\Delta t$.
Here, we denote the time steps
for the anion, cation, and flow velocity by $\Delta t_{\mathrm{a}}$,
$\Delta t_{\mathrm{c}}$, and $\Delta t_{\mathrm{u}}$, respectively.
The {\color{black}kinematic viscosity} is normally much
larger than the diffusion coefficients of the ion species,
{\color{black} i.e.,
the Schmidt number defined as $\mathrm{Sc}=\nu/D_m$ is 
large}.
Therefore the timescale of the flow is shorter than that of the diffusion
process, i.e., $\Delta t_{\mathrm{u}}\ll \Delta t_{\mathrm{a}},\,\Delta t_{\mathrm{c}}$.
We show in Fig.~\ref{fig02}
the iterative procedure 
for the case of $N_t\Delta t_{\mathrm{u}}=\Delta
t_{\mathrm{a}}=\Delta t_{\mathrm{c}}$,
which is the case discussed in Section~\ref{sec_result}
{\color{black} where $\mathrm{Sc}\sim 90$ and thus
$N_t=100$ is chosen.}
At each instance in $t$,
the electrical potential $\phi$ is obtained
as the limit at which $\tilde{t}\to \infty$.
Practically, however, the iteration with respect to $\tilde{t}$ is terminated
when the difference between two successive values of $\phi$
reaches a certain tolerance, typically $10^{-7}$~V,
at all of the lattice points.

%
%
%
\section{\label{sec_lubrication}Analytical model based on the
 lubrication approximation theory}

\begin{figure}[tb]
%
\begin{center}
\vspace*{0mm}
\includegraphics[scale=0.8]{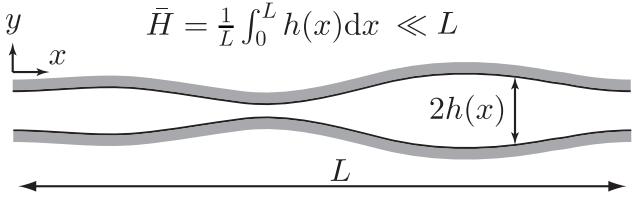}
\end{center}
%
\caption{Lubrication approximation.}
\label{fig03}
\end{figure}

In this section,
we outline the procedure to obtain an
analytical model for the electro-osmotic flow rate.
The method of analysis is 
based on the lubrication approximation theory,
which was first applied to the electro-osmotic flow 
in Ref.~\cite{Ghosal2002} for the case of a very thin electrical double layer,
where the Helmholtz--Smoluchowski approximation was assumed.
To simplify the set of model equations
described in Section~\ref{sec_problem} such that
the analytical method is applicable,
we assume the following:
(i) steady flow, (ii) small
convection effect in Eq.~\eqref{s2-np} (the third term on the right-hand
side) compared with
the other two terms,
(iii) small zeta potential (the electrical potential at the
interface) so that the Debye--H\"uckel approximation is valid, and 
(iv) larger length scale for the variation of the surface structure and
of the surface charge density along the channel
than the channel width, as depicted in Fig.~\ref{fig03}
($\bar{H}\ll L$; $\bar{H}=(1/L)\int_0^Lh(x)\mathrm{d}x$).
Assumption (iv) results in $\delta=\bar{H}/L\ll 1$ 
and this parameter $\delta$ is
used as a small parameter in the expansion analysis of
the lubrication approximation theory.

Before discussing the expansion analysis,
we introduce the following dimensionless variables:
\begin{align}
&x=\tilde{x}L,\quad
y=\tilde{y}\bar{H},\quad
\bm{u}=\tilde{\bm{u}}u_0,
\label{s4-nondim1}\\
&p=\tilde{p}\left(\dfrac{\mu u_0 L}{\bar{H}^2}\right),\quad
\phi_e=\tilde{\phi}_e\left(\dfrac{\bar{H}\sigma_0}{\varepsilon}\right),\quad
\phi_*=\tilde{\phi}_*\Delta\phi,
\label{s4-nondim2}\\
&h(x)=\tilde{h}(\tilde{x})\bar{H},\quad
\sigma(x)=\tilde{\sigma}(\tilde{x})\sigma_0,
\label{s4-nondim3}
\end{align}
{\color{black}
where  the reference flow velocity
$u_0$ and the inverse Debye length $\kappa$ are defined as
\begin{align}
u_0=\left(\dfrac{\kappa^2\bar{H}^3\sigma_0\Delta\phi}{\mu L}\right),
\quad
\kappa=\left(\dfrac{2C_0Fe}{\varepsilon k_{\mathrm{B}}T}\right)^{1/2},
\label{s4-kappa}
\end{align}
and} $\mu=\nu\rho_0$ is the viscosity of the electrolyte solution,
$C_0$ is the reference value of the ion concentration,
$\sigma_0$ is the reference value of the surface charge density,
and $\Delta \phi$ is the potential difference between
$x=0$ and $x=L$ (Fig.~\ref{fig01}).
Because the Debye--H\"uckel approximation is assumed
and the Poisson equation is hence linearized, the electrical potential
is split into the equilibrium potential $\phi_e$
and the potential describing the external electric field $\phi_*$.

Using the variables defined in Eqs.~\eqref{s4-nondim1} through
\eqref{s4-nondim3},
we transform the governing equations 
into a dimensionless form.
Here and in the following, the tilde attached to the dimensionless variables 
is dropped for simplicity, unless otherwise stated.
The set of equations then reads
\begin{align}
&\delta^2\dfrac{\pd^2\phi_*}{\pd x^2}+\dfrac{\pd^2\phi_*}{\pd y^2}=0,
\label{s4-poi1}\\
&\delta^2\dfrac{\pd^2\phi_e}{\pd x^2}+\dfrac{\pd^2\phi_e}{\pd
 y^2}=(\bar{H}\kappa)^2\phi_e,
\label{s4-poi2}\\
&0=-\dfrac{\pd p}{\pd x}+\delta^2\dfrac{\pd^2u_x}{\pd x^2}
+\dfrac{\pd^2u_x}{\pd y^2}+\phi_e\dfrac{\pd\phi_*}{\pd x}
+\gamma\delta\phi_e\dfrac{\pd\phi_e}{\pd x},
\label{s4-ns1}\\
&0=-\dfrac{1}{\delta}\dfrac{\pd p}{\pd y}+\delta^2\dfrac{\pd^2u_y}{\pd x^2}
+\dfrac{\pd^2u_y}{\pd y^2}+\dfrac{1}{\delta}\phi_e\dfrac{\pd\phi_*}{\pd y}
+\gamma\phi_e\dfrac{\pd\phi_e}{\pd y},
\label{s4-ns2}
\end{align}
where $\gamma=\sigma_0L/\varepsilon\Delta\phi$.
The boundary conditions at $y=\pm h(x)$ are
\begin{align}
&n_x\delta\dfrac{\pd\phi_*}{\pd x}+n_y\dfrac{\pd\phi_*}{\pd y}=0,
\label{s4-bc1}\\
&n_x\delta\dfrac{\pd\phi_e}{\pd x}+n_y\dfrac{\pd\phi_e}{\pd y}=-\sigma,
\label{s4-bc2}\\
&u_x=u_y=0,
\label{s4-bc3}
\end{align}
where the components of the normal vector at
the surfaces are expressed as 
$n_x=\pm \delta h'/(1+\delta^2h^{\prime 2})^{1/2}$ and 
$n_y=\mp 1/(1+\delta^2h^{\prime 2})^{1/2}$, with
$h'$ being the derivative of $h(x)$.
Here, the Nernst--Planck equation has been integrated
under assumptions (i) and (ii) described above to have the
Boltzmann distribution for the ion concentration, and 
has been further simplified under assumption (iii)
to be incorporated as in Eqs.~\eqref{s4-poi2} through \eqref{s4-ns2}.

To analyze the boundary-value problem described above,
we expand the variables with respect to the small parameter $\delta$:
\begin{align}
\phi_*=\phi^{(0)}_*+\delta \phi^{(1)}_*+\delta^2 \phi^{(2)}_*+\cdots,
\label{s4-expand1}\\
\phi_e=\phi^{(0)}_e+\delta \phi^{(1)}_e+\delta^2 \phi^{(2)}_e+\cdots,
\label{s4-expand2}\\
\bm{u}=\bm{u}^{(0)}+\delta \bm{u}^{(1)}+\delta^2 \bm{u}^{(2)}+\cdots,
\label{s4-expand3}\\
p=p^{(0)}+\delta p^{(1)}+\delta^2 p^{(2)}+\cdots.
\label{s4-expand4}
\end{align}
The goal of this analysis is to obtain the solution describing the
electro-osmotic flow for the limit of $\delta\to 0$, i.e., the solution $u_x^{(0)}$.
After substituting the expansions into Eqs.~\eqref{s4-poi1} through
\eqref{s4-ns2}, we equate the terms of the 
same power of $\delta$ to obtain the series of equations and
boundary conditions 
for the coefficients in the expansion, such as $\phi_*^{(n)}$ and
$\bm{u}^{(n)}$. The series of boundary-value problems are solved from the lowest order.

We first analyze 
the boundary-value problems for $\phi^{(n)}_*$
resulting from Eqs.~\eqref{s4-poi1} and \eqref{s4-bc1}.
The analysis up to the order of $\delta^2$
yields the following expression for $\phi_*^{(0)}$:
\begin{equation}
 \dfrac{\pd\phi_*^{(0)}}{\pd x}=\dfrac{1}{w_1h}, \quad
 \dfrac{\pd\phi_*^{(0)}}{\pd y}=0,
\label{s4-sol1}
\end{equation}
where $w_1$ is a quantity defined via
\begin{equation}
 w_k=\int_0^1\tilde{h}(\tilde{x})^{-k}\od \tilde{x}.
\label{s4-defw}
\end{equation}
In Eq.~\eqref{s4-defw}, the tilde 
explicitly shows that $\tilde{h}$ and $\tilde{x}$ are dimensionless.
Turning to the analysis of the boundary-value
problems derived from Eqs.~\eqref{s4-poi2} and \eqref{s4-bc2},
the solution $\phi^{(0)}_e$ is readily
obtained from the leading-order analysis as
\begin{equation}
 \phi^{(0)}_e=\dfrac{\sigma\cosh(\kappa \bar{H}
	y)}{\kappa\bar{H}\sinh(\kappa\bar{H}h)}.
\label{s4-sol2}
\end{equation}
We next analyze the leading-order problem derived from Eq.~\eqref{s4-ns2}.
From the result shown in Eq.~\eqref{s4-sol1}, we find that $p^{(0)}$ only depends on $x$:
\begin{equation}
 \dfrac{\pd p^{(0)}}{\pd y}=0.
\label{s4-sol3}
\end{equation}
The leading-order problem that results from Eqs.~\eqref{s4-ns1}
and \eqref{s4-bc3} is then considered.
Using all the results in Eqs.~\eqref{s4-sol1} through \eqref{s4-sol3},
we arrive at the following expression for $u_x^{(0)}$:
\begin{align}
& u_x^{(0)}=\dfrac{1}{2}\dfrac{\pd p^{(0)}}{\pd x}(y^2-h^2)\nonumber\\
& \hspace*{1.5cm}
+\dfrac{\sigma}{w_1h(\kappa\bar{H})^3}\left(\cosh(\kappa\bar{H}h)-\cosh(\kappa\bar{H}y)\right).
\label{s4-sol4}
\end{align}
The electro-osmotic flow rate is obtained by integrating this
expression over the channel width, i.e., $Q=2\int_0^{h}u_x\od y$.
However, the unknown variable $\pd p^{(0)}/\pd x$ is still included
in this expression.
The unknown variable is eliminated by using the constraint of the periodic boundary
condition for the pressure field, i.e., $\int_0^1(\pd p^{(0)}/\pd x)\od x=0$.
Then, we obtain the explicit formula for the electro-osmotic flow rate
in the limit of $\delta\to 0$ as follows:
\begin{equation}
 Q=\dfrac{2\sigma_0\Delta \phi}{\mu\kappa^2 L w_1 w_3}\int_0^1
\dfrac{\tilde{\sigma}(\tilde{x})}{\tilde{h}(\tilde{x})^3}
\left(\dfrac{\kappa\bar{H}}{\tanh(\kappa\bar{H}\tilde{h})}
-\dfrac{1}{\tilde{h}}\right)\od\tilde{x},
\label{s4-final}
\end{equation}
where $w_1$ and $w_3$ are dimensionless quantities defined in Eq.~\eqref{s4-defw}.
Here, the tilde is explicitly shown on the dimensionless variables $\tilde{x}$,
$\tilde{h}$, and $\tilde{\sigma}$
defined in Eqs.~\eqref{s4-nondim1} and \eqref{s4-nondim3},
whereas $Q$ denotes the dimensional volumetric flow rate per unit length in the $z$ direction.

%
%
%
\section{\label{sec_result}Results and discussion}

%
\subsection{\label{ssec_constant}Constant surface charge density}

\begin{figure}[tb]
%
\begin{center}
\vspace*{0mm}
\includegraphics[scale=0.8]{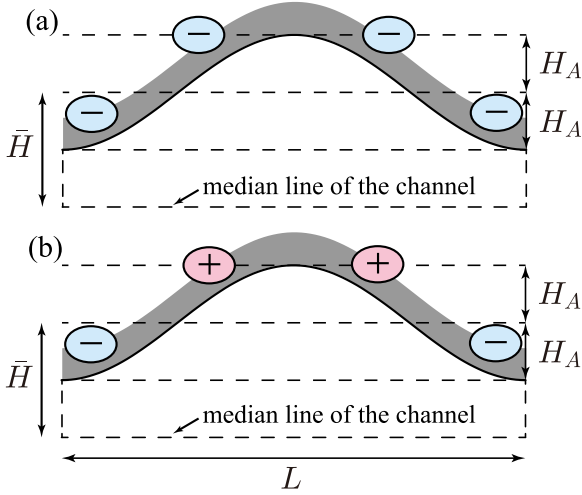}
\end{center}
%
\caption{Schematic diagram of the surface shape and the charge distribution.}
\label{fig04}
\end{figure}  

\begin{figure}[t]
\begin{center}
\vspace*{0mm}
\includegraphics[scale=0.65]{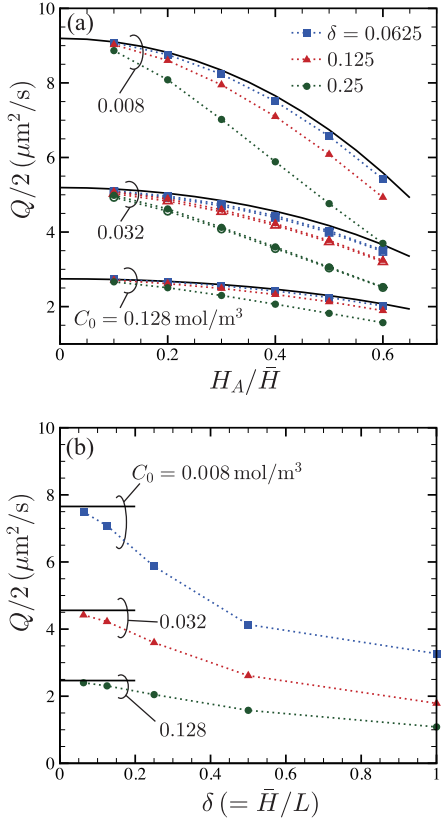}
\end{center}
%
\vspace*{-5mm}
\caption{
{\color{black}
Electro-osmotic flow rate in the channel with undulated
 walls
(a) as functions of $H_A/\bar{H}$ and (b) 
as functions of $\delta$ at $H_A/\bar{H}=0.4$.}
The surface charge density is constant ($\sigma=-2\times 10^{-5}$\,C/m$^2$).
The solid lines indicate the analytical model in the limit of $\delta\,(=\bar{H}/L)\to 0$.
The symbols indicate the results of the direct numerical simulation
by using the LBM in the case of finite values of $\delta$.}
\label{fig05}
\end{figure}

The electro-osmotic flow through
channels with the sinusoidal surfaces,
as depicted in Fig.~\ref{fig04}, is considered in the present study.
More precisely, the shape of the surfaces is defined as
\begin{equation}
 h(x)=\bar{H}-H_A\cos\left(\dfrac{2\pi x}{L}\right),
\label{s5-sin}
\end{equation}
where $H_A$ is the amplitude of the surface structure.
We first consider the case in which the surface charge is
distributed uniformly along the channel (Fig.~\ref{fig04}(a)).
We plot in Fig.~\ref{fig05} the flow rate for cases
in which the mean salt concentration $C_0$ is
$0.008$, $0.032$, and $0.128$\,mol/m$^3$,
as functions of the amplitude of the surface shape $H_A$
in panel (a) and
as functions of the geometrical parameter $\delta=\bar{H}/L$ in
panel (b).
The analytical results obtained from Eq.~\eqref{s4-final}
are indicated by the solid lines,
and the results of the numerical analysis obtained by using the LBM
are indicated by the symbols.
Whereas the analytical result
is for the limit 
$\delta=\bar{H}/L\to 0$, 
the LBM simulations are performed for finite
values of $\delta$. 
The following parameters are used to characterize the
electrolyte solution:
$\rho_0=1\times 10^3$\,kg/m$^3$, $\nu=0.889\times 10^{-6}$\,m$^2$/s,
$D_{\mathrm{a}}=D_{\mathrm{c}}=1\times 10^{-8}$\,m$^2$/s,
$-z_{\mathrm{a}}=z_{\mathrm{c}}=1$, and 
$\varepsilon=6.95\times 10^{-10}$\,C$^2$/Jm.
The temperature is $273$\ K, the strength of the external electric field is 
$\Delta\phi/L=1\times 10^4$\,V/m,
the constant surface charge density is 
$\sigma=-2\times 10^{-5}$\,C/m$^2$,
and the mean channel width is $\bar{H}=0.5$\,$\mu$m.
We note here that, with the above parameters,
the Debye length (or the thickness of the electrical double layer)
is comparable to the channel width, i.e.,
$\kappa\bar{H}=4.86$ for $C_0=0.008$, $\kappa\bar{H}=9.72$ for $C_0=0.032$, and 
$\kappa\bar{H}=19.4$ for $C_0=0.128$
(see Eq.~\eqref{s4-kappa}).

Since the Debye length $1/\kappa$ 
shortens with the increase of the mean salt concentration
and the electrical double layer becomes thin,
less volume of the fluid 
feels the driving force due to the 
electric field for larger salt concentrations.
The decrease of flow rate $Q$ 
is therefore observed in Fig.~\ref{fig05}(a), as the increase of salt
concentration $C_0$.
Generally, the flow rate decreases with the increase
of amplitude $H_A$; that is,
the electro-osmotic flow of the undulated surfaces
is small compared with that of
the straight channel ($H_A=0$) having the width $\bar{H}$.
This is consistent with the previous results,
where the surface roughness is found to decrease the 
electro-osmotic flow rate~\cite{WWC2007,BN2010,LWCR2010,MS2010}.

The analytical formula given in Eq.~\eqref{s4-final}
is the flow rate in the limit of $\delta\to 0$. 
In other words,
the frequency of the surface-shape variation
is infinitesimal,
because the surface shape is 
expressed in terms of the parameter $\delta$ as
$h(x)=\bar{H}-H_A\cos(2\pi \delta x/\bar{H})$.
Since the numerical data
that are obtained for the finite values of $\delta$
properly converge to the analytical results
with the decrease of $\delta$,
the appropriateness of analytical formula \eqref{s4-final} is confirmed.
{\color{black}The finite frequency effect
decreases the flow rate, and the influence is
larger for the larger values of amplitude $H_A$. 
We also plot in Fig.~\ref{fig05}(b) 
the dependency on the parameter $\delta=\bar{H}/L$ at $H_A/\bar{H}=0.4$.
The deviation from the analytical value exhibits
non-linear behavior, i.e., 
the flow rate quickly decreases for small values of $\delta$
and then gradually decays for large $\delta$,
which is consistent with the previous results, e.g., Fig.~3 of Ref.~\cite{MS2010}.}

Before concluding this subsection, we briefly mention the computational
system used in the the LBM.
The lattice points are placed uniformly with lattice spacing
$\Delta x/\bar{H}=0.01$
{\color{black}(note that the lattice spacing divided
by the characteristic length is often referred to as
the Knudsen number in the LBM.)}
For example,
in the case of $\delta=0.0625$, $1600$ lattice points are distributed
along the $x$ axis. 
We checked the sensitivity of the results on the size of the grid spacing
by comparing the results with those under a coarser lattice system.
The results with $\Delta x/\bar{H}=0.02$ are shown for the case of
$C_0=0.032$\,mol/m$^3$
in Fig.~\ref{fig05}(a) by the open symbols. Since the results of $\Delta x/\bar{H}=0.02$ are sufficiently close to
those of $\Delta x/\bar{H}=0.01$ (closed symbols), we safely use the
latter grid spacing for all computations in the present paper.
{\color{black}
The time step used for electrolyte flow is $\Delta
t_{\mathrm{u}}=0.0025$\,ns, and that for ion diffusion is $\Delta
t_{\mathrm{c}}=\Delta t_{\mathrm{a}}=0.25$\,ns.
We note again that the difference comes from the large Schmidt number
($\mathrm{Sc}\sim 90$), and $N_t=100$ is chosen in the present work (See
Fig.~\ref{fig02}).}

%
\subsection{\label{ssec_distributed}Inhomogeneously distributed surface charge}

Next, we consider the case in which the surface charge is distributed
along the channel, such that the net surface charge vanishes.
\begin{equation}
\int_0^L\sigma(x)\od x=0.
\label{s5-sc}
\end{equation}
In this case, regions
are charged positively and negatively, and
the driving forces in the opposite directions are exerted
if the external electric field is applied in the $x$ direction.
If the channel surfaces do not have the structure 
and the channel is straight, the driving forces in the opposite
directions cancel and no net flow occurs in the $x$ direction.
However, if the surfaces are structured and the
cross section varies along the channel, 
the symmetry of the forces can be broken
and one-way flow in the $x$ direction is expected.
We note here that the idea of making one-way flows by means of the
driving forces in the opposite directions has also been used
in designing the Knudsen pump for a rarefied gas~\cite{SWA1996,ADTY2007,ADM+2008,ATTY2010}.
In the Knudsen pump, the thermal transpiration flows, which occur in opposite
directions and are driven by the periodic temperature gradients,
are used to extract a one-way flow by introducing the structured channel
walls.
The application of 
structured surfaces to induce one-way electro-osmotic flows
was also proposed by Ajdari in Refs.~\cite{Ajdari1995,Ajdari1996},
under the assumption of infinitesimal thickness of the electrical double layer.
Here we demonstrate the possibility of one-way flow
in the case of the present setup, i.e., the 
thickness of the electrical double layer being comparable to the channel width.

\begin{figure}[t]
%
\begin{center}
\vspace*{0mm}
\includegraphics[scale=0.5]{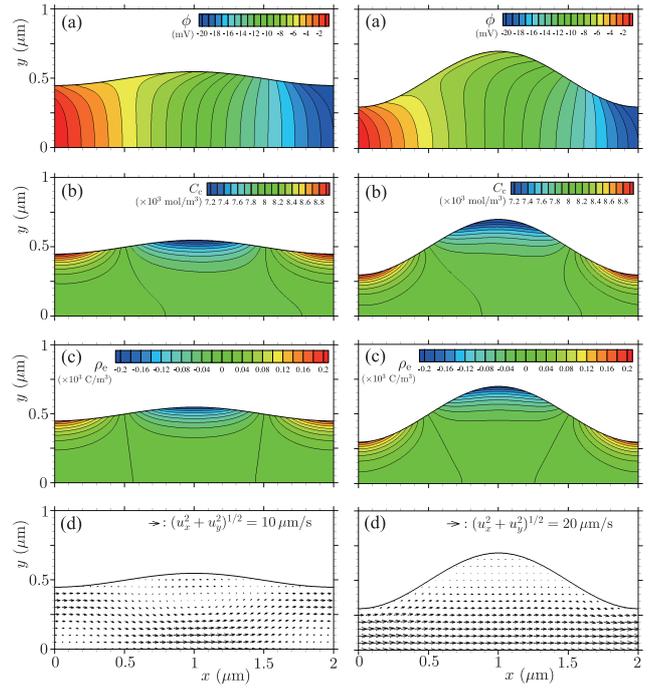}
\end{center}
%
\caption{{\color{black}
Two-dimensional plots of 
(a) electrical potential $\phi$, 
(b) cation distribution $C_{\mathrm{c}}$, 
(c) local charge density $\rho_{\mathrm{e}}$,
and (d) vector field $\bm{u}$,
in the channel with inhomogeneous surface charge distribution.
The average ion concentration is $C_0=0.008$\,mol/m$^3$,
and the amplitude of the surface shape is $H_A/\bar{H}=0.1$ (left panels),
or $H_A/\bar{H}=0.4$ (right panels).
}}
\label{fig06}
\end{figure}  

We consider the case in which the surface charge $\sigma(x)$ is distributed
along the channel as
\begin{equation}
\sigma(x)=\sigma_0\cos\left(\dfrac{2\pi x}{L}\right),
\label{s5-sigma}
 \end{equation}
where $\sigma_0=-2\times 10^{-5}$\,C/m$^2$. As schematically shown
in Fig.~\ref{fig04}(b),
the negatively charged region is narrower than 
the positively charged region,
and the density of the driving force
is larger in the negatively charged region.
Hence, if the electric field is applied in the $x$ direction,
one-way flow is expected to take place in the positive direction along the $x$ axis,
because the electrical double layer of the narrower region is positively charged.
In Fig.~\ref{fig06},
we show typical plots of electrical potential $\phi$, 
{\color{black} cation distribution $C_{\mathrm{c}}$,}
local charge density $\rho_{\mathrm{e}}$,
and flow velocity vector $\bm{u}$,
obtained by using the LBM.
The ion concentration is $C_0=0.008$\,mol/m$^3$,
and the geometrical parameters are $\bar{H}/L=0.25$,
{\color{black}
with $H_A/\bar{H}=0.1$ (left panels), or with $H_A/\bar{H}=0.4$ (right panels).}
The other parameters used are the same as those
used in the previous subsection. 
Since the electric field is induced by the potential bias of $\Delta \phi=-20$\,mV
as shown in Figs.~\ref{fig06}(a)
the driving forces act in the positively and negatively charged regions
shown in Figs.~\ref{fig06}(c).
{\color{black}
Note that the asymmetric distribution of 
ion shown in Figs.~\ref{fig06}(b) caused by the potential bias
is not captured by the Poisson--Boltzmann formulation,
because the Poisson--Boltzmann equation must be decoupled from
the external electric field.
}
The flow field
in the case of small amplitude ($H_A/\bar{H}=0.1$, {\color{black}the left panel of} Fig.~\ref{fig06}(d))
shows interference of the flows in the opposite directions
and a vortex is observed around $x=1$\,$\mu$m.
On the other hand,
in the case of large amplitude ($H_A/\bar{H}=0.4$, {\color{black}the right
panel of }Fig.~\ref{fig06}(d)),
the region where the driving force is exerted in the negative direction
is mostly hidden inside the wider region,
and strong one-way flow in the $x$ direction occurs in the middle part
of the channel.

\begin{figure}[t]
%
\begin{center}
\vspace*{-5mm}
\includegraphics[scale=0.7]{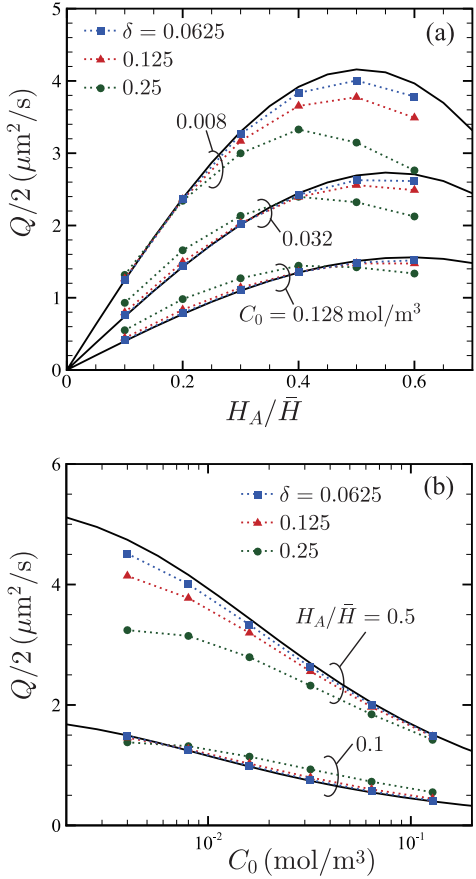}
\end{center}
%
\vspace*{-1mm}
\caption{\color{black}
Electro-osmotic flow rate in the channel with undulated
 walls (a) as functions of the amplitude $H_A/\bar{H}$,
and (b) as functions of the salt concentration $C_0$.
The surface charge is distributed inhomogeneously along the
 channel.
See the caption of Fig.~\ref{fig05}.
}
\label{fig07}
\end{figure}  

To investigate the 
dependency of the electro-osmotic flow rate on the geometrical parameter $H_A$,
we consider the flow rate $Q$ 
obtained from the LBM computations and that from the analytical formula
given by Eq.~\eqref{s4-final}
as functions of $H_A$, as shown in Fig.~\ref{fig07}(a).
In the case of the straight channel, i.e., $H_A=0$, the flow rate
is indeed zero, because the flows in the opposite directions cancel.
The analytical results (Eq.~\eqref{s4-final}) indicated by the solid line
show that the flow rate increases with the amplitude of the surface
shape, and it peaks around $H_A/\bar{H}=0.5$--$0.6$.
The LBM results also show the same tendency, but the
peaks shift slightly.
It is also noted that for the cases of $C_0=0.032$ and
$0.128$\,mol/m$^3$,
the effect of the finite value of $\delta$ raises the flow rate,
i.e., that the fast variation of the surface shape
and that of the surface charge density enlarge the electro-osmotic flow
in these parameter ranges.

{\color{black}
The analytical model~\eqref{s4-final} shows a
strong non-linear dependency on 
important quantities
such as dielectric constant $\varepsilon$, temperature $T$,
and salt concentration $C_0$,
all through the inverse Debye length $\kappa$ defined in Eq.~\eqref{s4-kappa}.
We plot in Fig.~\ref{fig07}(b) the 
electro-osmotic flow rate as functions of $C_0$  to examine this
dependency choosing $C_0$ as a controllable parameter.
The flow rate slowly decreases with increasing $C_0$ thus with increasing
$\kappa$ (note the horizontal axis is in log scale).
It is also observed that the effect of finite $\delta$ 
is significant for small values of $C_0$ with large amplitude $H_A/\bar{H}$.}

{\color{black}
The analysis of the problem considered in this section
confirmed the occurrence of the one-way flow, even in
the case of the zero net surface charge over the channel.
In the next stage of our study, we will evaluate
the possibility of the realization of one-way flow
experimentally, and we
will investigate applications of the present analysis methods to problems that include more complicated geometries, which 
we encounter in experimental setups such as porous media.
}

%
%
\section{\label{sec_summary}Conclusion}
In the present study, 
we numerically investigated the electro-osmotic flows
between two surfaces having a periodic structure
by using the lattice Boltzmann method (LBM),
and analytically by using the
lubrication approximation theory
for a moderate periodic structure. 
The numerical analysis is based on the
coupled LBM framework developed in Ref.~\cite{YKW2014},
where the Navier--Stokes {\color{black}equations} for the fluid flow,
the Nernst--Plank equation for the ion concentrations,
and the Poisson equation for the electric field are
simultaneously solved.
In the latter analysis, a
formula describing the electro-osmotic flow rate
induced in the microchannel, which has 
a width comparable to the thickness of the electrical
double layer,
was derived for the first time
for the case of structured surfaces
and inhomogeneous surface charge distributions.

In the analysis of the electro-osmotic flow through channels with the undulated surfaces
charged at a constant surface charge density,
the flow rate was
found to be suppressed by the large surface structure,
even for the infinitely slow variation of the surface shape.
The LBM results for the finite frequency of the surface-shape variation
showed that the fast variation of the surface shape 
further decreases the flow rate.
The analysis of the inhomogeneously distributed surface charge
confirmed the occurrence of the one-way flow, even under the
constraint of the zero net surface charge over the channel.
These results could pave the way for the design of novel devices
utilizing the electrokinetic flows in microchannels.



\section*{Acknowledgments}
The authors thank S. Iwai for computer assistance.
The present work was partially supported by the MEXT program ``Elements
Strategy Initiative to Form Core Research Center'' (since 2012).  (MEXT:
Ministry of Education, Culture, Sports, Science, and Technology, Japan.)






\begin{thebibliography}{10}
\expandafter\ifx\csname url\endcsname\relax
  \def\url#1{\texttt{#1}}\fi
\expandafter\ifx\csname urlprefix\endcsname\relax\def\urlprefix{URL }\fi
\expandafter\ifx\csname href\endcsname\relax
  \def\href#1#2{#2} \def\path#1{#1}\fi

\bibitem{Israelachvili2011}
J.~N. Israelachvili, Intermolecular and surface forces 3rd Edition, Academic
  press, 2011.

\bibitem{KBA2005}
G.~Karniadakis, A.~Beskok, N.~Aluru, Microflows and Nanoflows, Springer, 2005.

\bibitem{UTL+2006}
J.~P. Urbanski, T.~Thorsen, J.~A. Levitan, M.~Z. Bazant, Fast ac
  electro-osmotic micropumps with nonplanar electrodes, Appl. Phys. Lett. 89
  (2006) 143508.

\bibitem{YW2013}
Q.~Yuan, J.~Wu, Thermally biased ac electrokinetic pumping effect for
  lab-on-a-chip based delivery of biofluids, Biomed. Microdevices 15 (2013)
  125--133.

\bibitem{SPB+2013}
A.~Siria, P.~Poncharal, A.-L. Biance, R.~Fulcrand, X.~Blase, S.~T. Purcell,
  L.~Bocquet, Giant osmotic energy conversion measured in a single
  transmembrane boron nitride nanotube, Nature 494 (2013) 455--458.

\bibitem{CGS2007}
Z.~Chai, Z.~Guo, B.~C. Shi, Study of electro-osmotic flows in microchannels
  packed with variable porosity media via lattice {B}oltzmann method, J. Appl.
  Phys. 101~(10) (2007) 104913.

\bibitem{SHR2008}
R.~B. Schoch, J.~Han, P.~Renaud, Transport phenomena in nanofluidics, Rev. Mod.
  Phys. 80 (2008) 839--883.

\bibitem{BC2010B}
L.~Bocquet, E.~Charlaix, Nanofluidics, from bulk to interfaces, Chem. Soc. Rev.
  39 (2010) 1073--1095.

\bibitem{ZY2012}
C.~Zhao, C.~Yang, Advances in electrokinetics and their applications in
  micro/nano fluidics, Microfluid. Nanofluid. 13 (2012) 179--203.

\bibitem{DGC2013}
R.~Dey, T.~Ghonge, S.~Chakraborty, Steric-effect-induced alteration of thermal
  transport phenomenon for mixed electroosmotic and pressure driven flows
  through narrow confinements, Int. J. Heat Mass Transf. 56 (2013) 251--262.

\bibitem{ZRA2014}
J.~Zudrop, S.~Roller, P.~Asinari, Lattice {B}oltzmann scheme for electrolytes
  by an extended {M}axwell--{S}tefan approach, Phys. Rev. E 89 (2014) 053310.

\bibitem{YMK+2014}
H.~Yoshida, H.~Mizuno, T.~Kinjo, H.~Washizu, J.-L. Barrat, Molecular dynamics
  simulation of electrokinetic flow of an aqueous electrolyte solution in
  nanochannels, J. Chem. Phys. 140 (2014) 214701.

\bibitem{YMK+2014a}
H.~Yoshida, H.~Mizuno, T.~Kinjo, H.~Washizu, J.-L. Barrat, Generic transport
  coefficients of a confined electrolyte solution, Phys. Rev. E 90 (2014)
  052113.

\bibitem{MNM2014}
A.~Mehboudi, M.~Noruzitabar, M.~Mehboudi, Simulation of mixed
  electroosmotic/pressure-driven flows by utilizing dissipative particle
  dynamics, Microfluid. Nanofluid. 17 (2014) 199--215.

\bibitem{PPD+2014}
D.~V. Patil, K.~N. Premnath, D.~Desai, S.~Banerjee, Electrodeposition modeling
  using coupled phase-field and lattice {B}oltzmann approach, Int. J. of Mod.
  Phys. C 25 (2014) 1340018.

\bibitem{TS2006}
F.~Tessier, G.~W. Slater, Modulation of electroosmotic flow strength with
  end-grafted polymer chains, Macromolecules 39 (2006) 1250--1260.

\bibitem{WWC2007}
M.~Wang, J.~Wang, S.~Chen, Roughness and cavitations effects on electro-osmotic
  flows in rough microchannels using the lattice {P}oisson-{B}oltzmann methods,
  J. Comput. Phys. 226 (2007) 836--851.

\bibitem{WK2009}
M.~Wang, Q.~Kang, Electrokinetic transport in microchannels with random
  roughness, Anal. Chem. 81 (2009) 2953--2961.

\bibitem{XMS+2009}
Z.~Xia, R.~Mei, M.~Sheplak, Z.~H. Fan, Electroosmotically driven creeping flows
  in a wavy microchannel, Microfluid. Nanofluid. 6 (2009) 37--52.

\bibitem{BN2010}
S.~Bhattacharyya, A.~K. Nayak, Combined effect of surface roughness and
  heterogeneity of wall potential on electroosmosis in microfluidic/nanofuidic
  channels, J. Fluid. Eng. 132 (2010) 041103.

\bibitem{MS2010}
R.~J. Messinger, T.~M. Squires, Suppression of electro-osmotic flow by surface
  roughness, Phys. Rev. Lett. 105 (2010) 144503.

\bibitem{LWCR2010}
J.~Liu, M.~Wang, S.~Chen, M.~O. Robbins, Molecular simulations of
  electroosmotic flows in rough nanochannels, J. Comput. Phys. 229 (2010)
  7834--7847.

\bibitem{BB2015}
S.~Bhattacharyya, S.~Bera, Combined electroosmosis-pressure driven flow and
  mixing in a microchannel with surface heterogeneity, Appl. Math. Model. in
  press.
\newblock \href {http://dx.doi.org/doi:10.1016/j.apm.2014.12.050}
  {\path{doi:doi:10.1016/j.apm.2014.12.050}}.

\bibitem{YKW2014}
H.~Yoshida, T.~Kinjo, H.~Washizu, Coupled lattice {B}oltzmann method for
  simulating electrokinetic flows: a localized scheme for the {N}ernst--{P}lank
  model, Commun. Nonlinear Sci. Numer. Simulat. 19 (2014) 3570--3590.

\bibitem{Ghosal2002}
S.~Ghosal, Lubrication theory for electro-osmotic flow in a microfluidic
  channel of slowly varying cross-section and wall charge, J. Fluid Mech. 459
  (2002) 103--128.

\bibitem{NZ2012}
C.-O. Ng, Q.~Zhou, Electro-osmotic flow through a thin channel with gradually
  varying wall potential and hydrodynamic slippage, Fluid Dyn. Res. 44 (2012)
  055507.

\bibitem{NQ2014}
C.-O. Ng, C.~Qi, Electroosmotic flow of a power-law fluid in a non-uniform
  microchannel, J. Non-Newtonian Fluid Mech. 208 (2014) 118--125.

\bibitem{N1991}
J.~Newman, Electrochemical Systems, 2nd Edition, Prentice-Hall, Englewood
  Cliffs, NJ, 1991.

\bibitem{MZ1988}
G.~R. McNamara, G.~Zanetti, Use of the {B}oltzmann equation to simulate
  lattice-gas automata, Phys. Rev. Lett. 61 (1988) 2332.

\bibitem{QDL1992}
Y.~H. Qian, D.~d'Humi\`eres, P.~Lallemand, Lattice {BGK} models for
  navier--stokes equation, Europhys. Lett. 17 (1992) 479.

\bibitem{CD1998}
S.~Chen, G.~D. Doolen, Lattice {B}oltzmann method for fluid flows, Annu. Rev.
  Fluid Mech. 30 (1998) 329--364.

\bibitem{S2001}
S.~Succi, The lattice {B}oltzmann equation for fluid dynamics and beyond,
  Oxford Univ. Press, New York, 2001.

\bibitem{LL2000}
P.~Lallemand, L.-S. Luo, Theory of the lattice {B}oltzmann method:
  {D}ispersion, dissipation, isotropy, {G}alilean invariance, and stability,
  Phys. Rev. E 61 (2000) 6546--6562.

\bibitem{Hetal2002}
D.~d'Humi\`eres, I.~Ginzburg, M.~Krafczyk, P.~Lallemand, L.-S. Luo,
  Multiple-relaxation-time lattice {B}oltzmann models in three dimensions,
  Philos. Trans. R. Soc. Lond. A 360 (2002) 437--451.

\bibitem{AKO2003}
S.~Ansumali, I.~V. Karlin, H.~C. {\"O}ttinger, Minimal entropic kinetic models
  for hydrodynamics, Europhys. Lett. 63 (2003) 798.

\bibitem{GGK2006}
M.~Geier, A.~Greiner, J.~G. Korvink, Cascaded digital lattice {B}oltzmann
  automata for high {R}eynolds number flow, Phys. Rev. E 73 (2006) 066705.

\bibitem{Ginzburg2012}
I.~Ginzburg, Truncation errors, exact and heuristic stability analysis of
  two-relaxation-times lattice {B}oltzmann schemes for anisotropic
  advection-diffusion equation, Commun. Comput. Phys. 11 (2012) 1439.

\bibitem{GZS2008}
Z.~Guo, C.~Zheng, B.~C. Shi, Lattice {B}oltzmann equation with multiple
  effective relaxation times for gaseous microscale flow, Phys. Rev. E 77
  (2008) 036707.

\bibitem{YN2010}
H.~Yoshida, M.~Nagaoka, Multiple-relaxation-time lattice {B}oltzmann model for
  the convection and anisotropic diffusion equation, J. Comput. Phys. 229
  (2010) 7774--7795.

\bibitem{GH2013}
T.~Geb\"ack, A.~Heintz, A lattice {B}oltzmann method for the
  advection-diffusion equation with {N}eumann boundary conditions, Commun.
  Comput. Phys. 15 (2013) 487--505.

\bibitem{LMK2013}
L.~Li, R.~Mei, J.~F. Klausner, Boundary conditions for thermal lattice
  {B}oltzmann equation method, J. Comput. Phys. 237 (2013) 366--395.

\bibitem{HL1997C}
X.~He, L.-S. Luo, Lattice {B}oltzmann model for the incompressible
  {N}avier--{S}tokes equation, J. Stat. Phys. 88 (1997) 927--944.

\bibitem{QO1993}
Y.~H. Qian, S.~A. Orszag, Lattice {BGK} models for the {N}avier--{S}tokes
  equation: Nonlinear deviation in compressible regimes, Europhys. Lett. 21
  (1993) 255--259.

\bibitem{WK2010}
M.~Wang, Q.~Kang, Modeling electrokinetic flows in microchannels using coupled
  lattice {B}oltzmann methods, J. Comput. Phys. 229 (2010) 728--744.

\bibitem{CPF2004}
F.~Capuani, I.~Pagonabarraga, D.~Frenkel, Discrete solution of the
  electrokinetic equations, J. Chem. Phys. 121 (2004) 973--986.

\bibitem{PCF2005}
I.~Pagonabarraga, F.~Capuani, D.~Frenkel, Mesoscopic lattice modeling of
  electrokinetic phenomena, Comput. Phys. Commun. 169 (2005) 192--196.

\bibitem{GZS2005a}
Z.~Guo, T.~S. Zhao, Y.~Shi, A lattice {B}oltzmann algorithm for electro-osmotic
  flows in microfluidic devices, J. Chem. Phys. 122 (2005) 144907.

\bibitem{CS2007}
Z.~Chai, B.~Shi, Simulation of electro-osmotic flow in microchannel with
  lattice {B}oltzmann method, Phys. Lett. A 364 (2007) 183--188.

\bibitem{WWL2008}
J.~Wang, M.~Wang, Z.~Li, Lattice evolution solution for the nonlinear
  {P}oisson-{B}oltzmann equation in confined domains, Commun. Nonlinear Sci.
  Numer. Simulat. 13 (2008) 575--583.

\bibitem{Zhang2011}
J.~Zhang, Lattice {B}oltzmann method for microfluidics: models and
  applications, Microfluid. Nanofluid. 10 (2011) 1--28.

\bibitem{CS2008}
Z.~Chai, B.~C. Shi, A novel lattice {B}oltzmann model for the {P}oisson
  equation, Appl. Math. Model. 32 (2008) 2050--2058.

\bibitem{PPB2014}
D.~V. Patil, K.~N. Premnath, S.~Banerjee, Multigrid lattice {B}oltzmann method
  for accelerated solution of elliptic equations, J. Comput. Phys. 265 (2014)
  172--194.

\bibitem{SWA1996}
Y.~Sone, Y.~Waniguchi, K.~Aoki, One-way flow of a rarefied gas induced in a
  channel with a periodic temperature distribution, Phys. Fluids 8 (1996)
  2227--2235.

\bibitem{ADTY2007}
K.~Aoki, P.~Degond, S.~Takata, H.~Yoshida, Diffusion models for {K}nudsen
  compressors, Phys. Fluids 19 (2007) 117103.p1--p21.

\bibitem{ADM+2008}
K.~Aoki, P.~Degond, L.~Mieussens, S.~Takata, H.~Yoshida, A diffusion model for
  rarefied flows in curved channels, Multiscale Model. Simul. 6 (2008)
  1281--1316.

\bibitem{ATTY2010}
K.~Aoki, S.~Takata, E.~Tatsumi, H.~Yoshida, Rarefied gas flows through a curved
  channel: {A}pplication of a diffusion-type equation, Phys. Fluids 22 (2010)
  112001.

\bibitem{Ajdari1995}
A.~Ajdari, Electro-osmosis on inhomogeneously charged surfaces, Phys. Rev.
  Lett. 75 (1995) 755--758.

\bibitem{Ajdari1996}
A.~Ajdari, Generation of transverse fluid currents and forces by an electric
  field: Electro-osmosis on charge-modulated and undulated surfaces, Phys. Rev.
  E 53 (1996) 4996.

\end{thebibliography}

\end{document}